\documentclass[referee]{aa} 
\usepackage{graphicx}
\usepackage{txfonts}
\begin{document}
   \title{Photometric distances to young stars in the inner Galactic disk}

   \subtitle{I. The L = 314$^{o}$ direction\thanks{Based on observations collected at Las Campanas Observatory, Chile},
   \thanks{Catalogs are only available at the CDS via anonymous ftp to cdsarc.u-strasbg.fr (130.79.128.5) \     
     or via http://cdsarc.u-strasbg.fr/viz-bin/qcat?J/???/???/???}}

   \author{Giovanni  Carraro  
          \inst{1,2}
          }

   \institute{ESO, Alonso de Cordova 3107, 19001,
           Santiago de Chile, Chile\\
              \email{gcarraro@eso.org}
         \and
             Dipartimento di Astronomia, Universit\'a di Padova, Vicolo Osservatorio 3, I-35122,
             Padova, Italy\\
             \email{giovanni.carraro@unipd.it}
             }

   \date{Received May 1, 2011; accepted , 2011}

 
  \abstract
   {The spiral structure of the Milky Way is nowadays receiving renewed attention thanks to the combined 
   efforts of observational campaigns in different wavelength regimes, from the optical to the radio.}
   {We start in the paper the exploration of  a number of key sectors (line of sights) in the inner Milky Way, where
   the spiral structure is still poorly known.}
   {We search for density enhancements of young stars that might plausibly be associated with spiral structure. To this aim we collect
   sufficiently wide-field UBVI photometry to allow us to probe in statistical sense the distribution in reddening and distance
   of young stars in the field. The intensive usage of U-band photometry - although heavily demanding in terms of observational efforts - 
   ensures robust determination of reddening and hence distance for
   stars of spectral type earlier than A0, which are well-known spiral arm tracers, even though no spectroscopic information are available.
   The fields we use are large enough to include
   in most cases well-studied Galactic clusters, which we use as bench-marks to assess  the quality and standardisation of the data, 
   and to validate our method.}
   {We focus in this paper on the line of sight to the Galactic longitude l=  314$^o$, where previous surveys already detected
   H$_{\alpha}$  emitters at different standard of rest velocities, and hence distances. The difficulty, however, to translate
   velocity into distance make predictions on the spiral structure quite vague.  
   First of all, we made exhaustive tests to show that our
   data-set is in the standard system, and calibrated our method using the two open clusters NGC~5617 and Pismis~19 
   which happen to be in the field, 
   and for which we found estimates of the basic parameters in full agreement with the literature.  
   We then applied the method to the general field stars and detected signatures of three different groups of stars, evenly distributed
   across the field of view, 
   at 1.5$^{+0.5}_{-0.2}$, 2.5$^{+0.3}_{-0.5}$,  and 5.1$^{+1.5}_{-1.1}$ kpc, respectively. These distances are compatible with 
   the location of the nowadays commonly accepted description
   of the Carina-Sagittarius and Scutum-Crux arms, at heliocentric distance of $\sim$ 2 an 5 kpc, respectively. As a consequence,
   we consider these groups to be good candidates to trace the location of these two inner arms.}
   {In line with previous studies, this investigation demonstrates once again how powerful the use of U-band photometry 
   is to characterize ensembles of young stars, and make
   predictions on the spiral structure of the Milky Way.}


   \maketitle
%

\section{Introduction}
The study of the spiral structure of the Milky Way, for long time dormant,
is nowadays quite an active field of astronomical research (Efremov 2010, Grosbol et al. 2011).
New wide-area surveys in different wave-length domains have been conducted in
recent years, which allowed us to improve significantly our knowledge of the spiral
structure of the Galaxy in several Galactic zones.\\

\noindent
Among the various important contributions, we 
would like to remind here three recent fundamental investigations, which - interestingly enough-
employed very different techniques, and demonstrate how this field
of research is lively nowadays.\\
The first is the Green Bank Telescope HII Region Discovery Survey (Anderson et al. 2011),
which detected a large number of newly-discovered HII regions in the first Galactic quadrant,
delineating for the first time the distant outer (Norma Cygnus) arm in that quadrant.\\
The second is the detailed study of the third quadrant performed in optical and radio
by V\'azquez et al. (2008, and references therein), which traced for the first time the local
(Orion, or Local arm) all the way to the outer disk, and the outer (Norma-Cygnus) arm.
This latter study however did not find clear indications of the Perseus arm, which
on the other hand, is very well traced in the second quadrant.\\
The third, finally,  is the first-ever detection of a distant arm 
beyond the Outer Arm - the arm traced by Anderson et al. (2011)
in the first quadrant - using 21 cm surveys (Dame \& Thaddeus 2011).

  \begin{figure*}
   \centering
   \includegraphics[width=14cm]{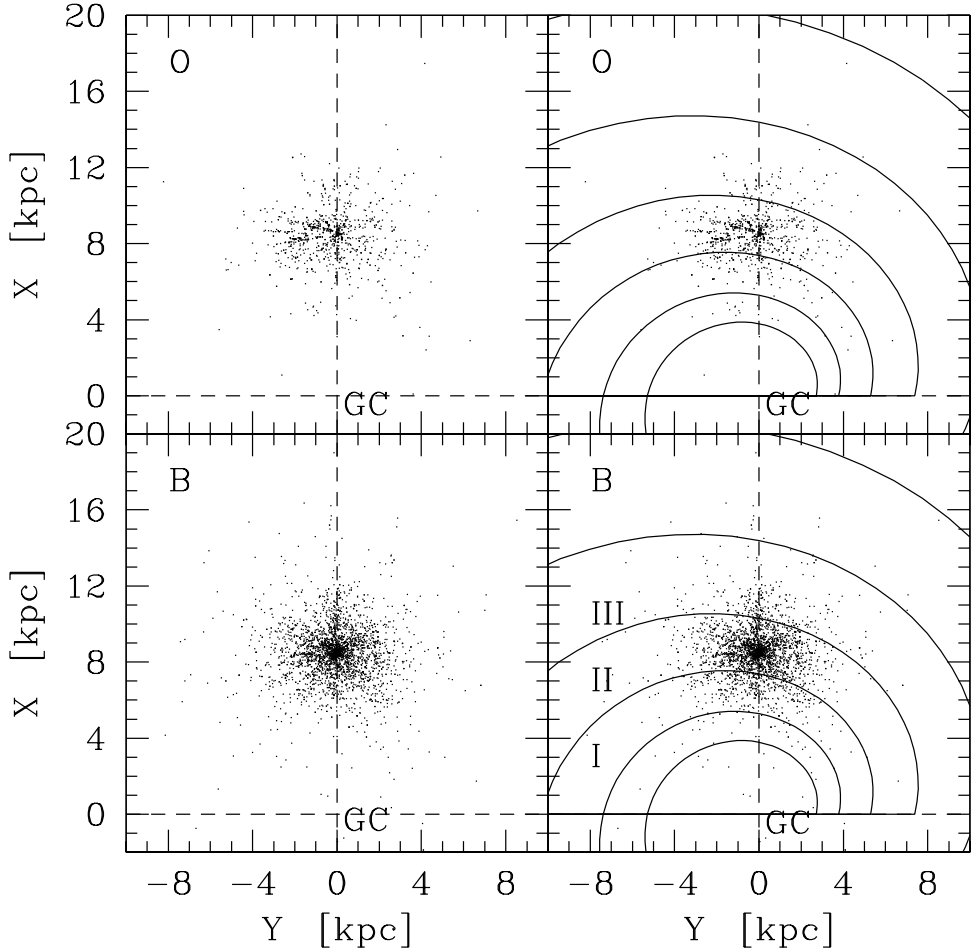}
   \caption{Distribution of OB type stars in the plane of the Milky Way as taken from Reed (2003) catalog.
            The lower panels show the distribution of B type stars alone (left panel), and with Vall\'ee
            2005 spiral arms super-posed (right panel). The same is shown in the upper panel, but for O
            type stars. The position of the Galactic Center (GC) is indicated, and the Sun is located
            at (0.0, 8.5). With the symbols I, II, and III in the bottom-right panel we indicate the Scutum-Crux, Carina-Sagittarius, and Perseus arm, respectively.}
    \end{figure*}

\noindent
Little efforts have been done in recent times to probe the spiral structure
in the fourth Galactic quadrant after the important discovery of a distant arm 
(most probably the outer, Norma-Cygnus arm) in HI by
McClure-Griffiths et al. (2004).
No indications have been found so far of the Perseus arm extension in this quadrant
- between l = 270$^o$ and 290$^o$ -
in spite of model predictions (Valle\'e 2005, Russeil et al. 2003).
The region around l=270$^o$ is complicated to study because of the presence of
the Vela molecular ridge close to the Sun which limits
significantly  optical and near infrared techniques (Carraro \& Costa 2010).
The line of sight to l=290$^o$ coincides with the tangent to the Carina-Sagittarium arm in the
fourth quadrant, making it again very difficult to detect features in optical or infrared 
beyond it (Shorlin et al. 2004, Carraro \& Costa 2009, Baume et al 2009).\\

\noindent
In a series of paper, Russeil (2003, and references therein)
carried out deep H$_{\alpha}$ survey of the inner Milky Way, and confirmed
the old picture that the inner disk is dominated by the Carina-Sagittarius arm
between l = 290$^{o}$ and 360$^{o}$ (Georgelin \& Georgelin 1976).
However, the difficulty to translate Local Standard of Rest (LST) velocities into distances makes 
it difficult to position with confidence features having velocities different from the expected one
for the conspicuous Carina arm. 
Only  in CO and HI, beyond Carina-Sagittarius, clear traces of Scutum-Crux - the molecular ring-
have been found in this Galactic sector (Dame \& Thaddeus 2011), which, again,
are tricky to interpret in terms of helio-centric distances.\\

\noindent
A more effective technique to measure distances to possible spiral features  
is the one extensively used by Carraro et al. (2005), Moitinho et al. (2006), V\'azquez et al. (2009) and Carraro et al. (2010),
which employs deep U$-$band photometry to estimate reddening of OB stars in the field or associated to star clusters,
and hence to constrain their distance.\\

\noindent
This technique is an old one, and was employed at the very beginning of spiral structure
research by Morgan et al.(1952). Historically, this method was motivated by the failure of star count techniques
to detect spiral features (Gingerich 1985) and by the evidence that in external spiral galaxies blue
stars and gas (typical population I objects) outline spiral arms.  Morgan et al. (1952) build up
the first compilation of  about 900 OB stars, whose distance were measured using UBV photometry,
and which delineated for the very first time the Orion and Perseus arm.
\noindent
The sudden rise of HI observations to trace the spiral arm of the Milky Way overtook the optical
campaigns for decades, until Georgelin and collaborators undertook in the seventies a survey of HII regions
and mapped several portion of the Galactic disk. Their results are summarized in Russeil (2003 and references therein)
and lead to the conclusion that our Galaxy is a four-arm spiral (Georgelin 1975).
\noindent
Attempts to use OB stars in the field or in young clusters were resumed as well in the seventies
with the UBV star cluster campaign  of the southern sky by
Moffat and Vogt (1972 and reference therein). This survey was however too shallow and could not be used
to trace spiral features beyond the Orion and Carina arm (say beyond about 2 kpc from the Sun).\\

\noindent
To summarize the state of the art in the field, we plot in Fig.~1 the projected distribution of OB stars
in the plane of the MW with and without alleged spiral arms (from Vall\'ee et al. 2005). The data are taken from the homogeneous
compilation performed by Reed (2003, and reference therein). This catalog\footnote{http://othello.alma.edu/~reed/filename} 
contains about 16,000 OB stars
for which spectro-photometric distances are known. In the lower panels we plot only B-type stars, while in the upper panel we restrict
ourself to O-type stars. Beside, to highlight possible spiral features, we only consider stars beyond
half an kpc from the Sun. The left panels do show  evidences of spiral arms only when O-type stars
are considered. The most striking evidence is a  blob of stars around the Sun and the many strikes
corresponding to best surveyed lines of sight.
In the upper panel only the tilted Orion spurs (Local arm) is evident. When alleged spiral arms are super-imposed,
a somewhat better picture appears, and it is possible to tentatively assign stars to more distant spiral
arms, like Perseus in the second Galactic quadrant and Carina in the first and fourth quadrants.
No clear association of stars to the Scutum-Crux arm is visible.\\

\noindent
In this paper, we apply this technique in a field (located at l = 314$^o$) where Russeil et al. (1998) found indications
of features with velocities larger than the ones expected for the Carina-sagittarius arm, 
and look for spiral features extending beyond that  arm, and hopefully in the Scutum-Crux one.
The technique is  the  same we employed in Carraro \& Costa (2009) and Baume et al. (2010).
Up to now, the only optical detection of this arm is reported by V\'azquez et al. (2005), where
a group of early type field stars compatible with the expected position of the arm have been found in the back-ground
of the open cluster Stock~16 at l= 306$^o$.1, b = 0.$^o$06.\\

\noindent
This line of sight (at l = 314$^o$) contains, as mentioned 
also by Russeil et al. (1998) the young star cluster NGC~5617,
and happens to contain also the old star cluster Pismis 19.
These two clusters are well known, and have been studied several times in the past.
NGC~5617 (Ahumada (2005), Kjeldsen \& Frandsen (1991)) is a young cluster $\sim$ 100 Myr old, while
Pismis 19 (Phelps et al. (1994), Carraro and Munari (2004), Piatti et al. (1998)) is much older, probably
around 1 Gyr.
The presence of these two clusters offers us the possibility first to check whether our photometry 
is in the standard system,
and second to validate our method to estimate the parameters for the young diffuse population, 
using star clusters as bench-marks.\\

\noindent
The present is a pilot study to introduce the method and illustrate the results for a 
representative line of sight where a possible detection of the Scrutum-Crux arm in optical 
is claimed for. A few other lines of sights 
have been considered in the first and fourth quadrant  
to specifically look for signatures of the presence of this arm in the first and fourth
quadrant. They are under analysis, and will be presented in a forth-coming paper. 
These lines of sight have been chosen following two criteria.
First, we considered directions where at least one well-known star cluster is present, 
to check the method and the results. Second, we used Scheghel et al. (1998) maps to control
that the reddening along these line of sight is relatively low to permit to see beyond
the Carina-Sagittarius arm.\\

\noindent
As a consequence, the present work is organized as follows. In Sect.~2 we describe how data were collected
and pre-reduced, discuss the photometric calibration, astrometry and completeness,
and compare our data with literature studies.
The reddening law toward l = 314$^o$ is discussed in Sect.~3, while Sect.~4 introduces the basics of the method
we use in this work to isolate groups of stars at different distances.
As an application of the method, we re-determine the fundamental parameters of NGC~5617 and Pismis~19 in Sects 5 and 6,
respectively.
The diffuse stellar population in the field is studied in Sect.~7, while in  Sect.~8 we discuss the spatial distribution
of the various groups we have detected in the field.
Our findings are finally summarized, and put
in the wider context of the spiral structure of the Milky Way in Sect.~9.

  \begin{figure}
   \centering
   \includegraphics[width=9cm]{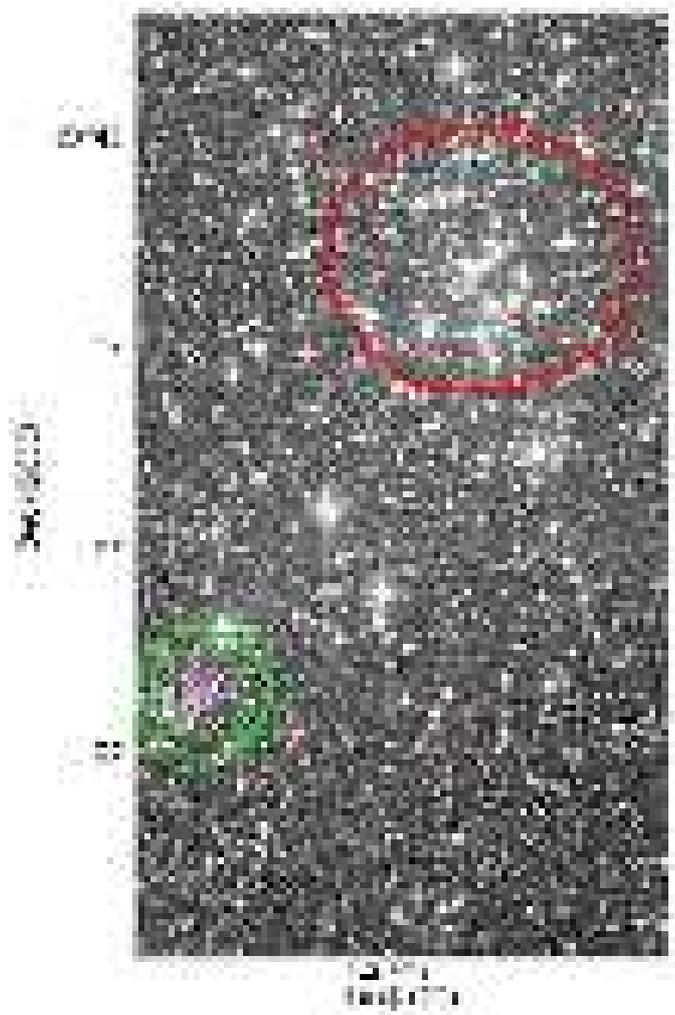}
   \caption{30 sec exposure in the V band showing the field studied in this paper.
            North is up, and East to the left. The field is 14.8 arcmin wide in RA,
            and 22.8 arcmin wide in DEC. Notice the presence of the old compact cluster Pismis 19
            in the south-east corner (green circle) , and the young diffuse cluster NGC~5617 in the north-west region
            (red circle)}
    \end{figure}

\section{Observations and data reduction}

The field under analysis was observed at Las Campanas Observatory (LCO) on the nights of June 26 and 29,
2006, as illustrated in Table~1, which lists useful details of the observations,
like filter coverage, airmass range and exposure time and sequences. 
We used the SITe$\#$3 CCD detector 
onboard the Swope 1.0m telescope\footnote{http://www.lco.cl/telescopes-information/henrietta-swope/}. 
With a pixel scale of 0.435 arcsec/pixel, this CCD allows
to cover 14.8 $\times$ 22.8 arcmin on sky. Both the nights were photometric with seeing
ranging from 0.9 to 1.4 arcsec. The field we covered is shown in Fig.~2, where a bias- and flat-field-
corrected image in the V band (30 secs) is shown.

To determine the transformation from our instrumental system to the standard Johnson-Kron-Cousins
system, and to correct for extinction, we observed stars in Landolt's areas 
PG~1047, PG~1323, SA~110, PG 1657, and MarkA (Landolt 1992)
multiple times and with different
air-masses ranging from $\sim1.05$ to $\sim2.0$, and covering quite a large color range 
-0.3 $\leq (B-V) \leq$ 1.7 mag. 
We secured night-dependent calibrations, which we then merged, after checking for
stability.

\begin{table}
\tabcolsep 0.1truecm
\caption{$UBVI$ photometric observations of target field and standard star.}
\begin{tabular}{lcccc}
\hline
\noalign{\smallskip}
Target& Date & Filter & Exposure (sec) & airmass (X)\\
\noalign{\smallskip}
\hline
\noalign{\smallskip}
SA~110       & 2006 Jun 26     & \textit{B} & 50, 60               &1.15$-$1.16\\
             &                 & \textsl{V} & 20, 30               &1.15$-$1.16\\
             &                 & \textsl{I} & 20, 30               &1.15$-$1.16\\
MarkA        & 2006 Jun 26     & \textit{B} & 90, 120              &1.07$-$1.38\\
             &                 & \textit{V} & 2x40                 &1.07$-$1.41\\
             &                 & \textit{I} & 2x40                 &1.07$-$1.43\\
PG~1323      & 2006 Jun 26     & \textit{B} & 30x2                 &1.11$-$1.81\\
             &                 & \textit{V} & 30x2, 90x2           &1.12$-$1.82\\
             &                 & \textit{I} & 30x2                 &1.13$-$1.85\\
Field        & 2006 Jun 26     & \textit{B} & 5, 60, 2x600         &1.08$-$1.25\\
             &                 & \textit{V} & 5, 30, 2x360         &1.07$-$1.22\\
             &                 & \textsl{I} & 5, 20x2, 2x300       &1.06$-$1.24\\
\hline
\hline

SA~110       & 2006 Jun 29     & \textit{U} & 180                  &1.15$-$1.16\\
             &                 & \textsl{B} & 120                  &1.15$-$1.16\\
             &                 & \textsl{V} & 40                   &1.15$-$1.16\\
             &                 & \textsl{I} & 40                   &1.15$-$1.16\\
MarkA        & 2006 Jun 29     & \textit{U} & 200                  &1.05$-$1.06\\
             &                 & \textit{B} & 120                  &1.05$-$1.06\\
             &                 & \textit{V} & 50                   &1.05$-$1.06\\
             &                 & \textit{I} & 50                   &1.05$-$1.06\\
PG~1323      & 2006 Jun 29     & \textit{U} & 150, 180             &1.11$-$1.81\\
             &                 & \textit{B} & 90, 100              &1.12$-$1.85\\
             &                 & \textit{V} & 60x2                 &1.12$-$1.83\\
             &                 & \textit{I} & 40x2                 &1.13$-$1.85\\
PG~1657      & 2006 Jun 29     & \textit{U} & 180, 240             &1.21$-$1.41\\
             &                 & \textit{B} & 90, 100              &1.22$-$1.45\\
             &                 & \textit{V} & 60x2                 &1.22$-$1.47\\
             &                 & \textit{I} & 40x2                 &1.23$-$1.45\\
Field        & 2006 Jun 29     & \textit{U} & 3x30, 3x90, 3x900    &1.08$-$1.26\\
             &                 & \textit{B} & 30, 60               &1.07$-$1.25\\
             &                 & \textsl{V} & 30, 60               &1.06$-$1.24\\
             &                 & \textit{I} & 30, 60               &1.07$-$1.25\\
\noalign{\smallskip}
\hline
\end{tabular}
\end{table}

\subsection{Basic photometric reduction}

Basic calibration of the CCD frames was done using IRAF\footnote{IRAF is distributed
by the National Optical Astronomy Observatory, which is operated by the Association
of Universities for Research in Astronomy, Inc., under cooperative agreement with
the National Science Foundation.} package CCDRED. For this purpose, zero exposure
frames and twilight sky flats were taken every night.  
All the frames were pre-reduced applying trimming, bias and flat-field
correction. Before flat-fielding, all frame were corrected for linearity,
following the recipe discussed in Hamuy et al. (2006).\\
Photometry was then performed
using the IRAF DAOPHOT/ALLSTAR and PHOTCAL packages. Instrumental magnitudes were extracted
following the point-spread function (PSF) method (Stetson 1987). A quadratic, spatially
variable, master PSF (PENNY function) was adopted, because of the large field
of view of the detector. Aperture corrections were then determined
making aperture photometry of a suitable number (typically 15 to 20) of bright, isolated,
stars in the field. These corrections were found to vary from 0.160 to 0.290 mag, depending
on the filter. The PSF photometry was finally aperture corrected, filter by filter.

\subsection{Photometric calibration}

After removing problematic stars, and stars having only a few observations in Landolt's
(1992) catalog, our photometric solution  for the run was extracted combining measures
from both nights- after checking they were stable and similar- yielding
a grand total of 83 measurements per filter, and turned out to be:\\

\noindent
$ U = u + (4.902\pm0.010) + (0.41\pm0.01) \times X + (0.129\pm0.020) \times (U-B)$ \\
$ B = b + (3.186\pm0.012) + (0.31\pm0.01) \times X + (0.057\pm0.008) \times (B-V)$ \\
$ V = v + (3.115\pm0.007) + (0.17\pm0.01) \times X - (0.057\pm0.011) \times (B-V)$ \\
$ I = i + (3.426\pm0.011) + (0.07\pm0.01) \times X + (0.091\pm0.012) \times (V-I)$ \\

\noindent
where $X$ indicates the airmass.\\
The final {\it r.m.s} of the fitting in this case was 0.030, 0.020, 0.013, and 0.013 
in $U$, $B$, $V$ and $I$, respectively.\\

\noindent
Global photometric errors were estimated using the scheme developed by Patat \& Carraro
(2001, Appendix A1), which takes into account the errors resulting from the PSF fitting
procedure (i.e., from ALLSTAR), and the calibration errors (corresponding to the zero point,
color terms, and extinction errors). In Fig.~3 we present our global photometric errors
in $V$, $(B-V)$, $(U-B)$, and $(V-I)$ plotted as a function of $V$ magnitude. Quick
inspection shows that stars brighter than $V \approx 20$ mag have errors lower than
$\sim0.05$~mag in magnitude and lower than $\sim0.10$~mag in $(B-V)$ and $(V-I)$. Higher
errors, as expected, are seen in $(U-B)$.\\

   \begin{figure}
   \centering
   \includegraphics[width=10cm]{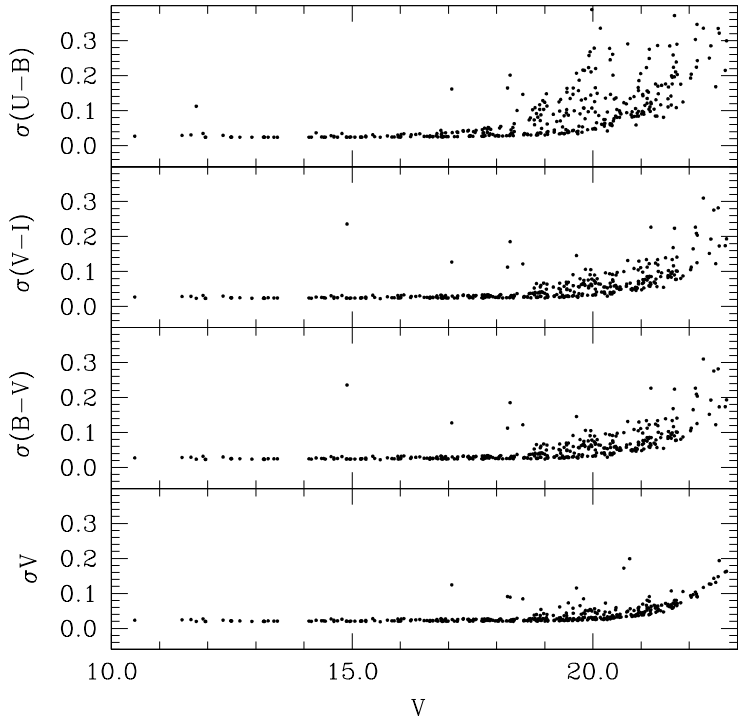}
   \caption{Trend of global photometric errors in magnitude and colors as a function of V magnitude.
   See text for details.}
    \end{figure}

\noindent
The final catalogue contains 3003 \textit{UBVI} and 13965 \textit{VI} entries.

\subsection{Completeness and astrometry}

Completeness corrections were determined by running artificial star experiments
on the data. Basically, we created several artificial images by adding artificial stars
to the original frames. These stars were added at random positions, and had the same
color and luminosity distribution of the true sample. To avoid generating overcrowding,
in each experiment we added up to 20\% of the original number of stars. Depending on
the frame, between 1000 and 5000 stars were added. In this way we have estimated that the
completeness level of our photometry is better than 90\% down to $V = 20.5$.\\

The optical catalog was then cross-correlated with 2MASS, which resulted in a final catalog
including \textit{UBVI} and \textit{JHK$_{s}$} magnitudes. As a by-product, 
pixel (i.e., detector) coordinates
were converted to RA and DEC for J2000.0 equinox, thus providing 2MASS-based astrometry, useful
for {\it e.g.} spectroscopic follow-up.
An excerpt of the optical photometri table used in this investigation is illustrated in Table~2.\\

\subsection{Comparison with previous photometry}
As mentioned in the Introduction, the region under investigation contains two Galactic clusters, which have been
studied several times in the past. This fact offer us the opportunity of assessing the quality of our
data-set and verify it is in the same system. Among the various possibility we choose Piatti et al. (1998)
to compare our VI photometry with, since they overlap in the Pismis 19 area. 
As for UBV, the only study which we can compare with is Kjeldsen and Frandsen (1991), which overlaps
with the present study in the area of the star cluster NGC~5617.
Although both these studies cover significantly smaller areas than the present study, still the number
of stars in common is statistically useful  for such a comparison.\\

\begin{table*}
\tabcolsep 0.3truecm
\caption{An excerpt of the optical photometric table exploited in this paper. 
The full version is posted at the CDS website. ID indicates
the numbering.}
\begin{tabular}{lcccccccccc}
ID & RA(2000.0) & DEC(2000.0) & V & $\sigma_V$ & (U-B) & $\sigma_{(U-B)}$ & (B-V) & $\sigma_{(B-V)}$ & (V-I) & $\sigma_{(V-I)}$\\    
\hline
 & deg & deg & & & & & & &\\
\hline
   1&  217.7315494&  -60.7259275&   15.732&    0.023&    0.377&    0.033&    0.709&    0.040&    0.963&    0.131\\
   2&  217.7335214&  -60.8564802&   17.822&    0.032&    0.696&    0.079&    0.977&    0.098&    1.497&    0.092\\
   3&  217.7320373&  -60.7914271&   17.715&    0.042&    1.611&    0.161&    1.633&    0.176&    2.141&    0.055\\
   4&  217.7336258&  -60.9069196&   17.802&    0.034&    0.947&    0.140&    1.689&    0.179&    1.867&    0.037\\
   5&  217.7311313&  -60.7492316&   17.912&    0.038&    1.418&    0.108&    1.153&    0.122&    1.333&    0.044\\
   6&  217.7287133&  -60.6497375&   15.557&    0.020&    0.525&    0.030&    0.539&    0.037&    0.672&    0.028\\
   7&  217.7296698&  -60.7161973&   18.972&    0.030&    1.182&    0.163&    1.223&    0.080&    1.508&    0.037\\
   8&  217.7326328&  -60.9264304&   18.020&    0.033&    1.013&    0.091&    1.549&    0.050&    1.995&    0.039\\
   9&  217.7330031&  -60.9553195&   13.990&    0.036&    2.195&    0.052&    2.332&    0.056&    2.517&    0.050\\
  10&  217.7321778&  -60.9455497&   16.459&    0.025&    0.304&    0.032&    0.753&    0.040&    0.958&    0.029\\
\noalign{\smallskip}
\hline
\end{tabular}
\end{table*}

\noindent
As for UBV, we found 129 stars in common with Kjeldsen and Frandsen (1991). The comparison is shown in Fig.~4,
in the sense of this study minus Kjeldsen and Frandsen (1991).
From this figure one can readily see that the two photometric data-sets are in the same system. However, while V and (B-V)
compare nicely, the (U-B) color, while in the same system, show quite a significant scatter.
However, most of the scatter in the (U-B) comes from stars fainter than V $\approx$ 17 mag, close to the 
limiting magnitude of Kjeldsen and Frandsen (1991) photometry, where photometric errors are much larger
than in the present study. 
As a confirmation, such an  increase in the scatter is also visibile
in the V and (B-V) comparisons at the same magnitude level, where the errors affecting our photometry are
lower than 0.02 mag (see Fig~2).
If we restrict the comparison for U-B to stars brighter than V $\approx$ 17, $\sigma$ drops to 0.05 mag.
We therefore conclude that the two studies are fully compatible down to  V $\approx$ 17. Below
this values, the scatter increases because of the increasing errors in the faint end of 
Kjeldsen and Frandsen (1991) photometry.\\

   \begin{figure}
   \centering
   \includegraphics[width=10cm]{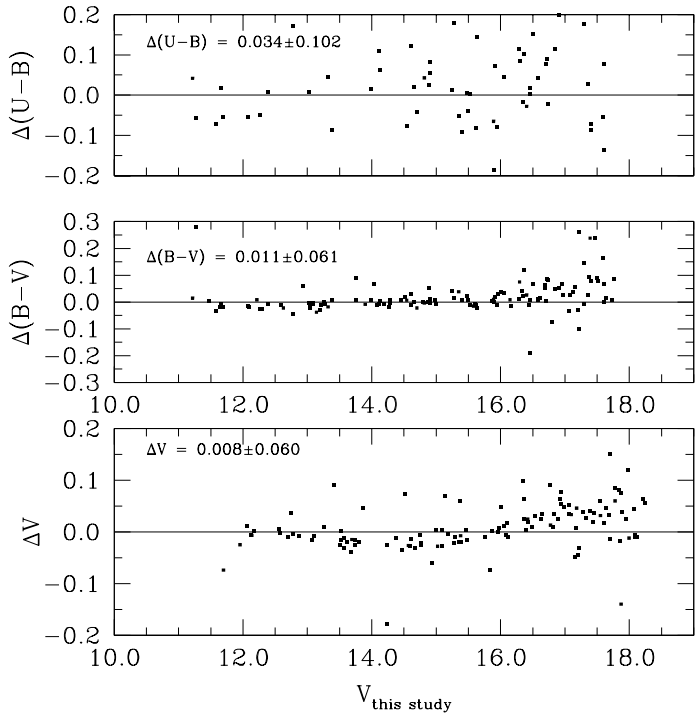}
   \caption{Comparison of our UBV photometry with Kjeldsen and Frandsen (1991) in the region of the open
    cluster NGC~5617.}
    \end{figure}

   \begin{figure}
   \centering
   \includegraphics[width=10cm]{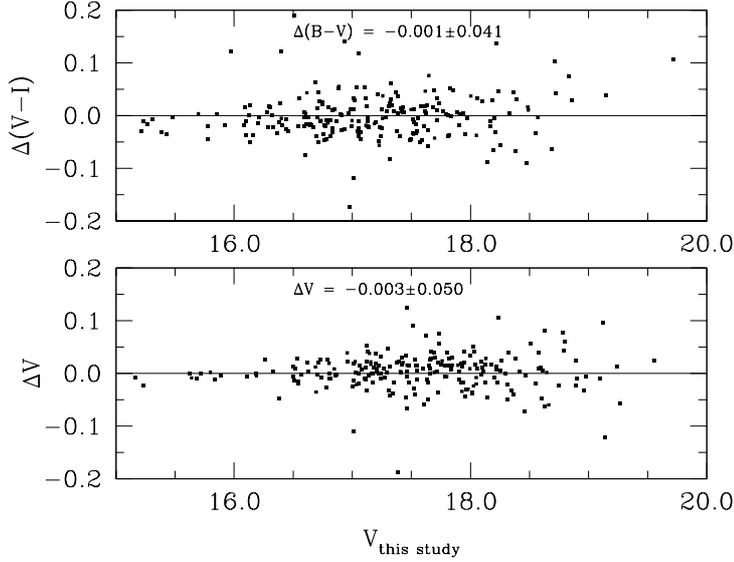}
   \caption{Comparison of our VI photometry with Piatti et al. (1998) in the region of the open
    cluster Pismis~19. }
    \end{figure}

\noindent
In the case of the VI photometry, we found 228 stars in common with  Piatti et al. (1998),
and the comparison is shown in Fig.~5, in the sense of this study minus Piatti et al. (1998). 
Again, the two studies compare nicely, which confirms
that our wide-field photometry is in the standard system.

\section{The reddening law toward l=314}
A basic requirement before analyzing our photometric material is to investigate the 
nature of reddening in this Galactic line of sight.
The reddening law is described by the ratio of total-to-selective absorption
R$_V$= $\frac{A_V}{E(B-V)}$, and the typical value in the Galaxy is about 3.1, with exceptions
in star-forming regions. 
Deviations from the standard value normally stand out in the
$(V-I)$ vs. $(B-V)$ color-color diagram, which for the region under study is
shown in Fig.~6. If the reddening law is normal, in this diagram stars follow - in other words, distribute with the same mean slope of- 
the reddening free relation, which for this color combination is $\frac{E(V-I)}{E(B-V)} = 1.244$ (Dean et al. 1978).
As reddening
free relation relation we adopt the Zero Age Main Sequence (ZAMS) from Schmidt-Kaler (1982),
which is drawn as a dashed red line for dwarf stars (Luminosity class (LC)  V) , 
and as a dashed blue line for giant stars (LC III).
The two long  arrows in the plot indicate the reddening vectors
for a normal reddening law ($R_V$= 3.1), and for an anomalous one ($R_V$=4.0), to guide the eye.
By inspecting Fig.~6 we conclude that the reddening law towards l = 314$^{o}$ does not show any evident
deviation,  and therefore
in the following discussion we will adopt $R_V$= $\frac{A_V}{E(B-V)} = 3.1$ and, in turn,
$\frac{E(U-B)}{E(B-V)} = 0.72$.

   \begin{figure}
   \centering
   \includegraphics[width=10cm]{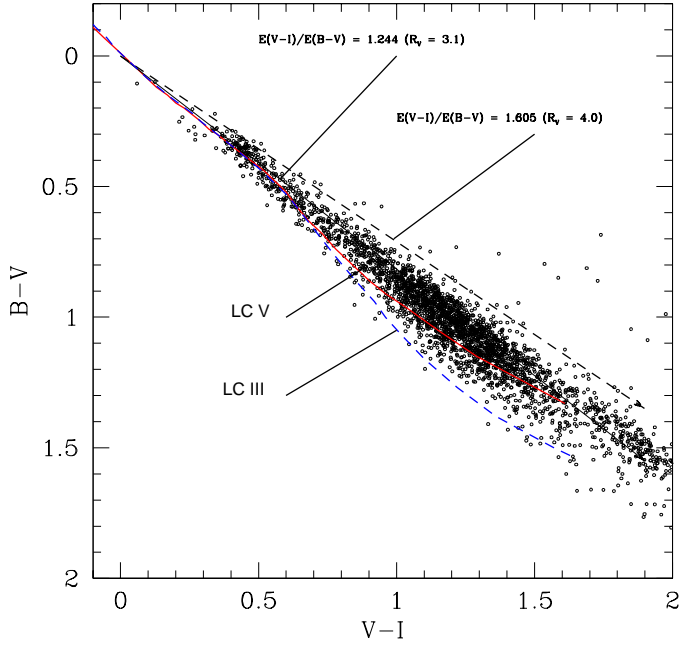}
   \caption{$(V-I)$ vs. $(B-V)$ color-color diagram for stars in our field with $UBVI$ photometry.
    The reddening vectors for the normal extinction 
    law  (R$_V$=3.1 and an anomlaous one (R$_V$=4.0) are plotted with arrows. The
    the red dashed line represents the reddening-free Schmidt-Kaler (1982) Zero Age Main
    Sequence relations for dwarf (LC V), while the blue is for giant stars (LC III).}
    \end{figure}

\section{The method}
Having shown that our data is in the standard system, and the reddening law looks normal,
we are now ready to use UBV photometry to measure young stars's reddening and distance.
The foundations of the method we are going to use are described in full 
details in Johnson (1965) and Straizys (1991).
We remind the reader that this purely photometric method, as described in the introduction,
has been succesfully used in the past to detect groups of field stars having
common reddening and distance in the background
of Galactic clusters (see, {\it e.g.}, Carraro et al. 2005, 2010;  V\'azquez et al. 2005, 2008;
Pandey et al. 2006, to cite a few example), both in the third and in the fourth
Galactic quadrant.\\

\noindent
Briefly, the heart of the method is based on the simultaneous inspection of the various photometric
diagrams to search for stars with common properties (reddening and distance).\\

\noindent
In the $(B-V)$ vs. $(U-B)$ color-color diagram (CCD), the position of a star only
depends on it reddening. 
To determine reddening, spectral type and photometric distance we then proceed
as follows.
First we derive intrinsic colors using the two relationships (from Straizys 1991),
which are valid if the reddening law is normal, as we have shown above:

\begin{equation}
E(U-B) = 0.72 \times E(B-V) + 0.05 \times E(B-V)^{2} ,
\end{equation}

\noindent
and

\begin{equation}
(U-B)_0 = 3.69 \times (B-V)_0 + 0.03.
\end{equation}

\noindent
The intrinsic color (B-V)$_0$ is the positive root of the second order equation one
derive combines the above expressions.
Intrinsic colors  ((U-B)$_0$ and (B-V)$_0$) are then directly 
correlated to spectral type, as compiled for instance in
Schmidt-Kaler (1982). The solution
of the equations above therefore allows us to encounter stars having spectral types
earlier than A0.5. For these stars we then know the absolute magnitude M$_{V}$ (again from the
Schmidt-Kaler 1982 compilation)
and, from  the apparent extinction-corrected magnitude V$_{0}$, we finally infer the photometric distance.\\
Errors in distances are then computed using a series of equations, as  follows:\\

\begin{equation}
\Delta (Dist)  = ln(10) \times Dist \times \Delta [log(Dist)];
\end{equation}

\begin{equation}
\Delta [log(Dist)] = \frac{1}{5} \times \Delta V + \Delta (M_V) + \Delta (A_V)];
\end{equation}

\begin{equation}
\Delta (M_V) = 0;
\end{equation}

\begin{equation}
\Delta (A_V) = 3.1 \times \Delta (B-V);
\end{equation}

\noindent
where $\Delta$ (V) and $\Delta(B-V)$ directly comes from photometry; finally:\\

\begin{equation}
\Delta (Dist) = ln(10) \times Dist \times \frac{1}{5} \times [ \Delta V + 3.1\times \Delta (B-V) ]
\end{equation}

\noindent
As anticipated above, this method is effective only for stars having spectral-type as late as A0V ( Straizys 1991), since
a unique reddening solution cannot evidently be determined for spectral types beyond
the A0V knee in the ZAMS. This is less than a problem for the purpose of this study, since
stars with spectral type earlier than A is exactly what we are looking for as spiral features' candidates.\\

\noindent
To caution the reader, we stress here that the distances we are obtaining are purely photometric
and are subject in most cases to large uncertianties, which depend not only on photometric errors,
but mostly on spectral type mis-classification. This is especially the case here since
we do not have spectroscopic confirmations for the spectral types of the stars we consider
as early type stars solely according to their position in the CCD. 
To partially cope with these limitations
we first check our method against well-known star clusters in the same field 
to see if we can recover their fundamental parameters with reasonably small uncertainties.\\

\noindent
In general, photometric distances will be anyway of inferior precision than, {\it eg.}, trigonometric parallaxes
(see Reid et al. 2009, and references therein), which, however, are 
unfortunately available only for a small number of lines of sight. 
When comparing different distance sources, Reid et al. (2009) emphasize that kinematic
distances  (derived mostly from CO and HI radio observations) can be  larger than trigonometric parallaxes
by factors greater than 2.0, especially in in the first and fourth
quadrants, because of distance ambiguities and the existence of sizeable noncircular motions. 
However, for the few lines of sight they considers - mostly towards Perseus and Sagittarius-
the position of star forming regions derived with the trigonometric parallax method do
not deviate significanly from the position inferred from spectro-photometric distances. Masers
located in the Perseus arms in the Galactic latitude range 115$^o$-189$^o$ are at virtually the same
heliocentric-distance range (2.1-2.6 kpc) as most young open clusters associated to Perseus,
in the same Galactic sector.
Consistency between trigonometric parallaxes and photometric distances have also been found by the same
group in the case of W3(OH) (Hachisuka et al. 2006), in the second Galactic quadrant.

\section{Checking the method: basic parameters of  NGC~5617}
The basic parameters of NGC~5617 have been measured in the past by Ahumada (2005) and
Kjeldsen and Frandsen (1991), and the two studies basically agree.
In fact,  Ahumada (2005) estimated a reddening E(B-V) = 0.54$\pm$0.09, and a distance modulus 
(m-M)$_o$ =11.53$\pm$0.40,
while Kjeldsen and Frandsen (1991) obtained E(B-V) = 0.48$\pm$0.02 and (m-M)$_o$ =11.55$\pm$0.20.
The results from  Kjeldsen and Frandsen (1991) are however to be considered more reliable,
since they are based on UBV photometry, while Ahumada (2005) estimates the reddening
by comparing the stars distribution in the color magnitude diagram with theoretical
isochrones.\\
\noindent
We are going to re-determine these parameters here
as a cross-check of our method. To this aim we isolate on a spatial basis NGC~5617 most probable
member as those stars which lie within 5 arcmin from the cluster center (Dias et al. 2002).
The selection is shown in the map in Fig.~1, where NGC~5617 is indicated as a 
red circular concentration  centered at RA= 217$^0$.436, Dec = -60$^0$.7187, J2000.0.

We constructed the $(B-V)$ vs. $(U-B)$ color-color diagram for these stars, which we show 
in Fig.~7. The  reddening-free Schmidt-Kaler (1982) Zero Age Main
Sequence relations for dwarf stars is drawn as a solid black line. The same relation, but shifted along the
reddening vector to fit the distribution of blue stars in NGC~5617 is drawn in blue.
The reddening vector is indicated in the plot with an arrow. To guide the eye, we indicate the
location of some spectral type stars in the reddening-free ZAMS, and move them
along the reddening vector to reach the reddened ZAMS which best fits NGC~5617 data..\\

\noindent
From this exercise we can derive the following conclusions:

\begin{itemize}
\item the reddening E(B-V) results to be 0.45$\pm$0.05, where the uncertainty
depends mostly on the width of NGC~5716 sequence.  It is difficult to say 
whether this broadness is caused by differential reddening, since photometric errors
($\sim$ 0.04 in U-B) account almost completely for that;
\item the reddening we derive is therefore in perfect agreement with previous studies;
\item NGC~5617 harbors stars with spectral type as early as B5, which implies an age
around 70$\pm$10 million years (Marigo et al. 2008).
\end{itemize}

   \begin{figure}
   \centering
   \includegraphics[width=10cm]{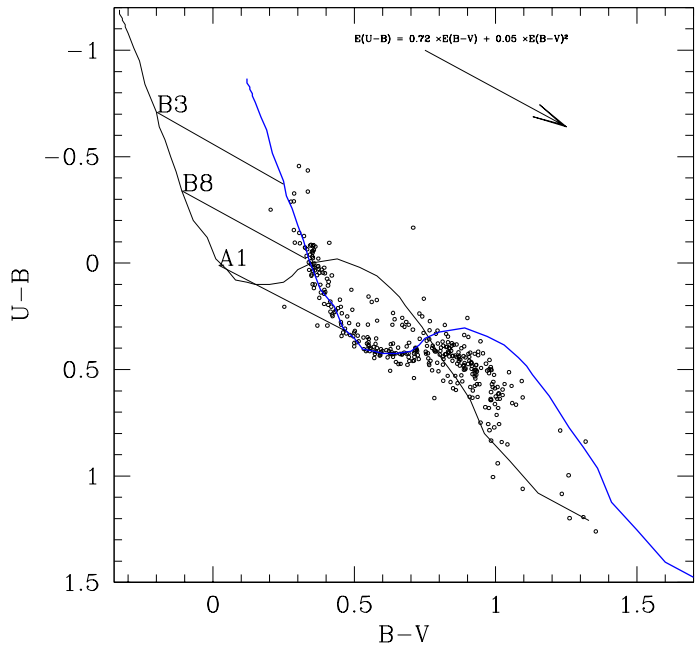}
   \caption{$(B-V)$ vs. $(U-B)$ color-color diagram for NGC~5617 stars within 5.0 arcmin from the cluster center. The solid lines
   are empirical ZAMS for no reddening (black line) and for E(B-V)=0.45 (blue line). The reddening
  vector is shown as an arrow in the top-right corner. The position of a few spectral types 
  is indicated to guide the eye.}
  \end{figure}

   \begin{figure}
   \centering
   \includegraphics[width=9cm]{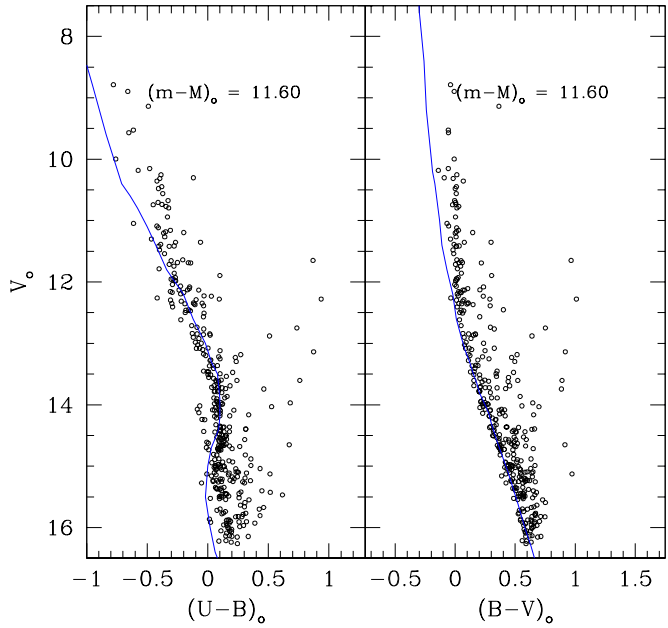}
   \caption{Reddening corrected color magnitude diagrams for NGC~5617 stars within 5.0 arcmin from the cluster center. The blue lines
   are empirical ZAMS vertically shifted for the value of the distance modulus indicated in both
   panels.}
  \end{figure}

   \begin{figure}
   \centering
   \includegraphics[width=10cm]{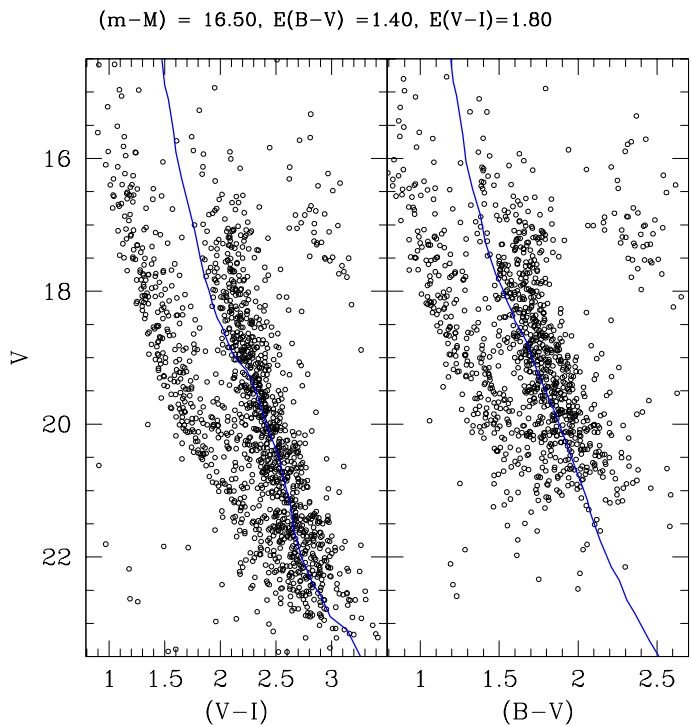}
   \caption{Color-magnitude diagrams for Pismis~19stars within 5.0 arcmin from the cluster center. 
   In the left panel the V vs (V-I) CMD is shown,
   while in the right panel we show the V vs B-V CMD. The blue lines are empirical ZAMS adjusted to
   fit the star distribution using the parameters indicated in the top of the figure. Notice in both
   the figure the vertical blue sequence of field stars in fron of Pismis ~19.}
  \end{figure}

\noindent
We are now in good position to estimate NGC~5617 heliocentric distance, and to
this aim we build up the reddening-corrected color magnitude diagram in (U-B)$_0$ vs V$_0$
and (B-V)$_0$ vs V$_0$. While the latter is the most widely used, the former has the great
advantage that the MS is tilted, allowing for a more reliable comparison
with empirical ZAMS.
This comparison is shown in Fig.~8 for (U-B) vs V (left panel), and (B-V) vs V (right panel).
Quite a good fit is reached shifting the ZAMS in vertical direction by (m-M)$_o$=11.60$\pm$0.15,
in both the diagrams. 
The uncertainty depends - again- on the width of the MS.\\

This yields an heliocentric distance to the cluster of 2.08$\pm$0.20 kpc, in fine agreement
with previous studies.
We note that the turn off point (TO) is located at V$_o$ $\sim$ 12.00$\pm$0.10 and hence at M$_{V}$=
0.40$\pm$0.10, which is consistent with an age around 70$\pm$10 million years (Marigo et al. 2008).

\noindent
At such an age, we expect NGC~5617 to be still very close to its birth-place.

   \begin{figure*}
   \centering
   \includegraphics[width=15cm]{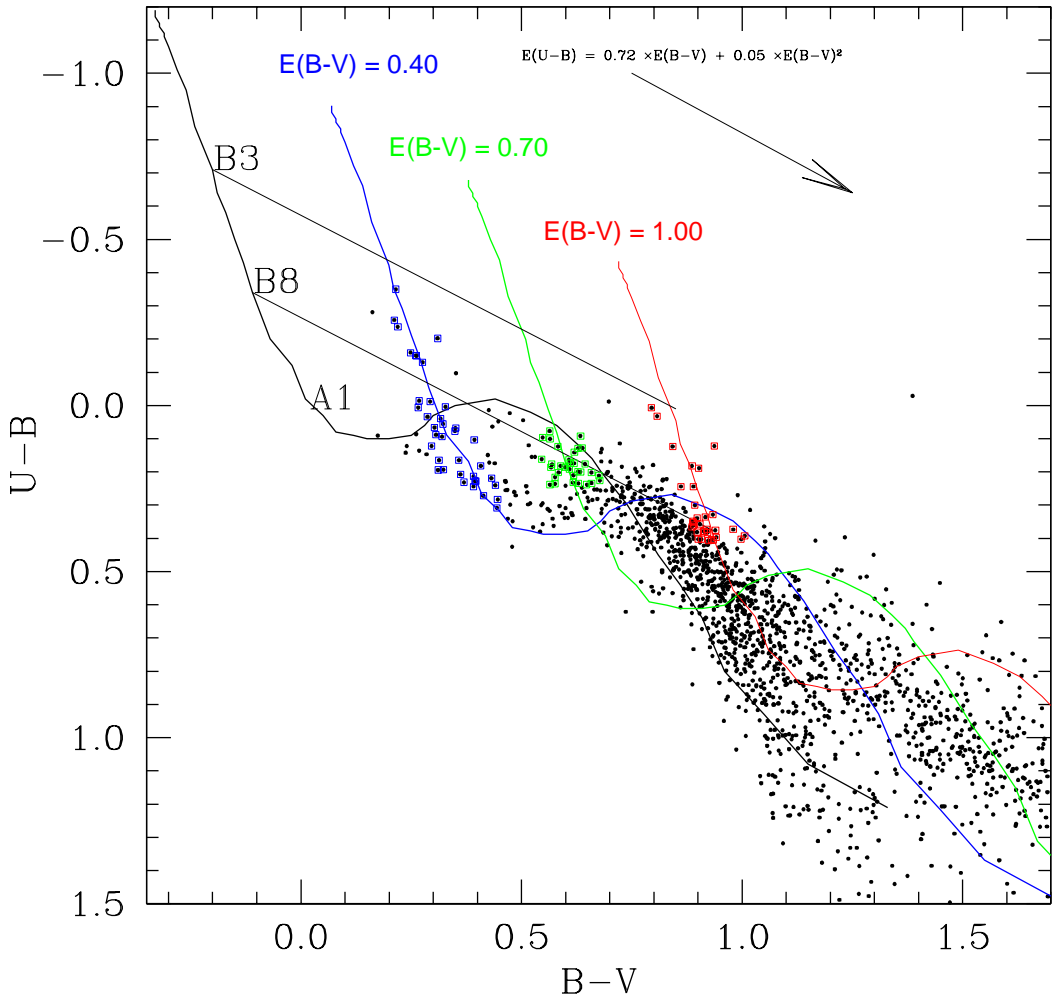}
   \caption{$(B-V)$ vs. $(U-B)$ color-color diagram for field stars. The solid lines
   are empirical ZAMS for no reddening (black line) and for E(B-V)=0.40 (blue line)
  0.70 (green line),  and 1.00 (red line). The reddening
  vector is shown as an arrow in the top-right corner. The position of a few spectral types 
  is indicated to guide the eye. Stars crowding along the various ZAMS are depicted
   with empty squares, colored in the same way as the corresponding ZAMS. See text for more
   details.}
  \end{figure*}

\section{Checking the method: basic parameters of Pismis~19}
The other cluster in the field, Pismis~19, has a radius of 3 arcmin (Dias et al. 2002).
Its stars are enclosed  in green circle in Fig.~1 centered at RA= 217$^0$.679, Dec = -60$^0$.889, J2000.0. 
The cluster is known to be heavily
reddened and of intermediate-age, and therefore we cannot rely on U photometry
to infer reddening and distance. We therefore make use of BVI photometry and fit to
the data an empirical ZAMS. 
The exercise is illustrated in Fig.~9, where the V vs (V-I) CMD is shown in the left panel,
and the V vs B-V CMD is shown in the right panel.
Only stars within the cluster radius are used.
First of all, we notice that our photometry gets almost 2 magnitudes deeper than previous
studies, which allows a more solid comparison with empirical sequences, since
the magnitude baseline is larger.
Beside, we would like to emphasize the presence of the vertical sequence of stars to the blue
of Pismis~19, which is composed by field stars located between us and the clusters -interlopers-,
and not to NGC~5617 (Phelps et al. 1994), which is about 15 arcmin away.\\

\noindent
We made use of the reddening-free Schmidt-Kaler (1982) Zero Age Main
Sequence relations for dwarf to fit the blue side of the MS, and the fit
yields the values indicated in the top of the plot.
The uncertainties on these parameters can be estimated by eye and amount to 0.2 for the reddening
and to 0.3 at least for the distance modulus.
These values are in very nice agreement with previous studies, and confirm
that Pismis~19 is indeed an highly obscured cluster. Its heliocentric distance is estimated
to be 2.5$\pm$0.5 kpc, somewhat further than NGC~5617.\\

\noindent
The presence of a clump of red giant stars at V$\sim$ 17 and B-V $\sim$ 2.4 suggests an intermediate age
for Pismis~19, and makes this cluster not useful for tracing spiral features. Therefore
providing a solid estimate of the age goes beyond the scope of this investigation.

\section{The young, diffuse, stellar population in the field}
The usage of our UBVI photometry - which is in the standard system - and the CCD and CMDs
confirmed previous literature results in the case of NGC~5617 and Pismis~19. 
We can therefore extend here the same technique to field stars,
with the aim of searching for common reddening stars in the field, which can indicate the presence of spiral
features at the same distance of NGC~5617 or even more distant, as found for instance in the field
of Shorlin~1 by Carraro \& Costa (2009), in the field of Danks~1 and 2 by Baume et al. (2010), 
and in the background of Stock~16 (V\'azquez et al. 2005).\\

\noindent
The color-color diagram of field stars is shown in Fig.~10. This has been constructed masking
the region of the young cluster NGC~5617 only, since no young blue stars in the field of Pismis~19 
actually belong to it.  To help the reader to follow our arguments, mainly based on visual inspection
of the diagram, we overplot the zero-reddening ZAMS, and three additional ZAMS, shifted according to
the reddening E(B-V) indicated on top of them. They have been drawn to highlight the location of common
reddening sequences. These sequences are made of early type stars shifted by interstellar reddening
and not falling in the crowded region of later than A0 type stars.
One cannot expect these sequences to be very tight, since they are made of field stars with different
spatial locations (see Carraro \& Costa (2009) for additional details).\\
\noindent

\noindent
The most prominent and obvious  sequence is the one through which we over-plot the E(B-V)= 0.40 (blue line) ZAMS. This is made 
of stars as early as B4.\\

\noindent
Two less conspicuous sequences are at average reddening of E(B-V) = 0.70 (green line) and 1.00 (red line), respectively.
While there might be even more reddened sequence, we consider them not very reliable, since they are at the limit of 
U-band sensitivity.\\

\noindent
We used the {\it Q-method} (Strayzis 1991) to extract stars having reddening around these three values, and counted
47, 41 and 29 stars close to the ZAMS reddened by about 0.40, 0.70 and 1.00 mag, respectively. For these sequences
we extimate a  mean reddening of  E(B-V)= 0.37$\pm$0.06, 0.76$\pm$0.05, and  1.05$\pm$0.09 mag, respectively.\\
The stars belonging to each common reddening group are indicated in the CCD with empty boxes, 
and with the same color as the ZAMS over-imposed for their mean reddening value.\\

\noindent
To appreciate the discreteness in distance of these three sequences we 
make use of the variable extinction diagram in Fig~11.
The variable extinction diagram usage is fully described in, {\it e.g.},
Johnson (1965). Briefly, it has been widely employed in the past to establish the reddening law in different
regions of the Milky Way. This reddening law, represented by the parameter $R_V$,  is the slope
in a diagram of apparent distance moduli (V-$M_V$) versus reddening (E(B-V)), and has been
succesfully used when large range of reddening  (namely variable extinction) was detected along the line of sight.
The usage of this diagram to estimate $R_V$ is possible when stars' spectral types are available
from spectroscopy, which allow to estimate their absolute magnitude independently from photometry.
In this study, we make use of the variable extinction diagram to show the reddening distribution
of the discrete groups we identified along the line of sight and visually appreciate their
distance distribution and spread, since we already know from the (V-I) vs (B-V) TCD that the reddening law
is normal. The dispersion in the X-axis of Fig~11
is the one derived using the  {\it Q-method} above, while the  dispersion in the Y-axis is graphically
shown with histograms using the same color coding as in Fig~10.\\
The three distributions
peaks at (V-M$_V$)= 12.10$\pm$0.40, 14.20$\pm$0.30, and 16.80$\pm$0.80.\\

\noindent
Once corrected for extinction, these distance moduli imply distances of 
1.5$^{+0.5}_{-0.2}$, 2.5$^{+0.3}_{-0.5}$,  and 5.1$^{+1.5}_{-1.1}$ kpc, respectively.\\

\noindent
We notice that the first two groups bracket NGC~5617 in distance. According to Bronfman et al. (2000)
the Carina-Sagittarius arm in this Galactic sector  is centered at about 2 kpc and has a width of about 1 kpc, which would
imply that the two groups are encompassing in heliocentric distance the arm, and NGC~5617 just falls inside
the arm.\\
The third group is much more scattered, mostly because of the significant uncertainty and spread in 
reddening. In spite of that, it is clearly  tracing  a group of young stars located beyond Carina-Sagittarius.

   \begin{figure}
   \centering
   \includegraphics[width=9cm]{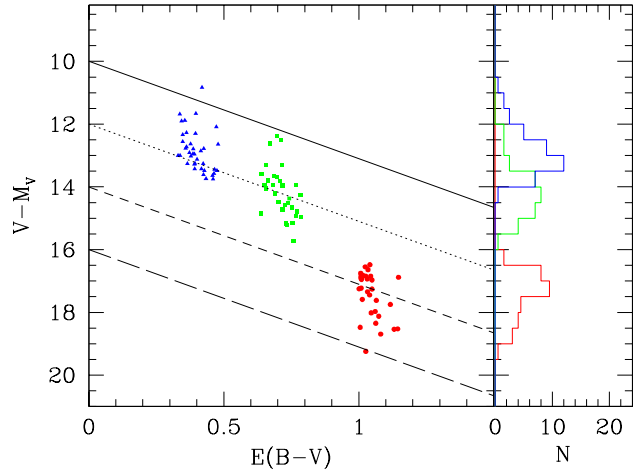}
   \caption{Variable extinction diagram for the groups of stars identified in the color-color
   diagram in Fig.~10. The color coding is the same as in Fig.~10.}
  \end{figure}

   \begin{figure}
   \centering
   \includegraphics[width=9cm]{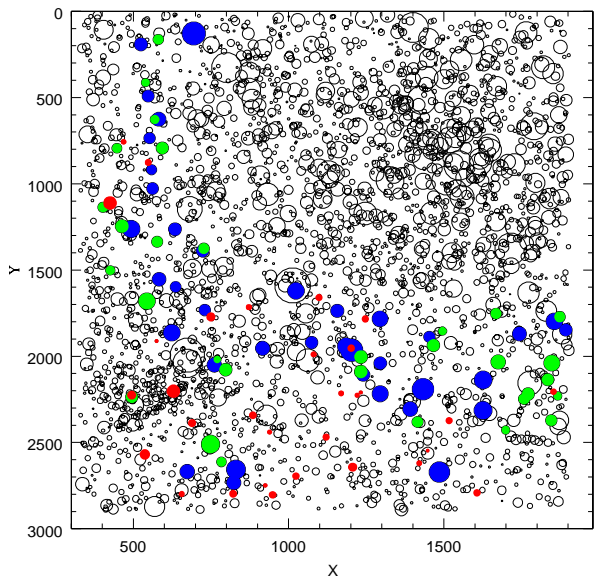}
   \caption{Spatial distribution of the stars belonging to the three common reddening groups
    we detected in the field of view. Color coding is the same as in Fig.~10.}
  \end{figure}

\section{Spatial distribution of B type stars in the field}
In this Section we are going to study the spatial distrution of the three groups
we identified in the CCD and described in previous Section. The underlying motivation is to investigate
whether the stars belonging to them are evenly distributed across the field, or are
grouped together. In the latter case, this would mean these bright stars are
part of some  cluster/association. \\

\noindent
The situation is illustrated in Fig.~12.
The stars in the region  under study are plotted with empty circles whose size is 
proportional to their magnitude. The star clusters NGC~5617 and Pismis~19 are clearly visible.
We then used the same colors as in Fig~10 to over-plot the young stars belonging to the three
different groups we indentified in the field, and which fall outside NGC~5617 area.\\
 
\noindent
Interestingly enough, they are evenly distributed across the field, and do not show any significant
clustering. We exclude therefore that these B-type stars are part of an overlooked
stellar cluster, but are most probably genuine field stars.

\section{Discussion and Conclusions}
In this study we have presented and discussed UBVI photometry in a field centered
at l=314, b=-0.6, with the aim of searching for young star candidates  to be part of the inner
Galaxy spiral arms.
In the same field of view two star clusters are present, NGC~5617 and Pismis~19, for which
we estimate fundamental parameters, and found they are in fine agreement with previous studies.
This make us confident that our photometry is in the standard system, and the method of analysis
correct.\\

\noindent
By analysing the TCD of field stars in the line of view, we detected three common reddening
groups, for which we provide mean reddening and distance.
The three group have reddening 
E(B-V)= 0.37$\pm$0.06, 0.76$\pm$0.05, and  1.05$\pm$0.09 mag, respectively, and are located at
distances of  1.5$^{+0.5}_{-0.2}$, 2.5$^{+0.3}_{-0.5}$,  and 5.1$^{+1.5}_{-1.1}$ kpc from the Sun.
We also showed that stars belonging to the three groups are evenly distributed across
the field of view, which means they do not belong to any over-looked star cluster.\\

\noindent
We now discuss the importance of theses groups as candidate spiral arm tracers. \\
Russeil et al. (1998) in the context of the deep H$_{\alpha}$ survey of the Milky Way
studied the same line of sight at l = 314$^o$. They found signatures of three different
radial velocity structures, which imply the presence of emitters at three different distances.
The lower velocity group, with LSR velocity V$_{lsr}$ $\sim$ 2-5 km/sec
corresponds to local, solar vicinity emission, with no traces of any condensation.\\
We do no find any hint for young stars so close to the Sun in our field,
most probably because the field of view is not sufficiently large to detect features that close
in a statistically significant way.\\

   \begin{figure}
   \centering
   \includegraphics[width=9cm]{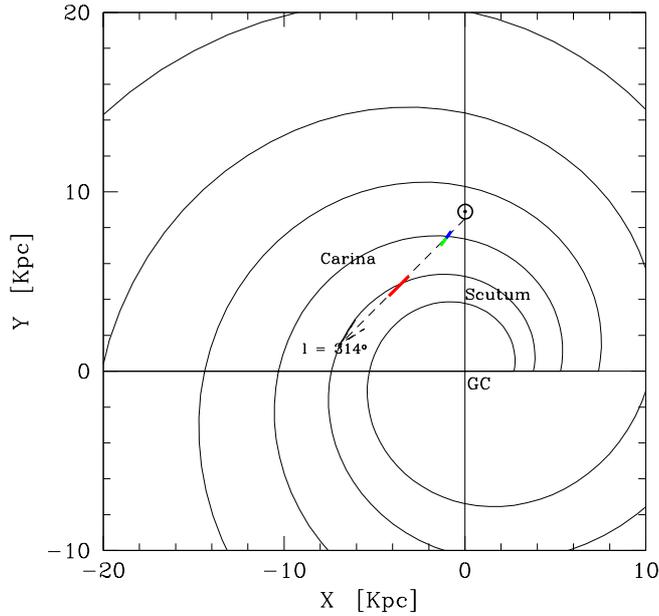}
   \caption{Vall\'ee (2005) realization of the Milky Way spiral structure. The position of the Sun is
   indicated with the usual symbol, while GC refers to the Galactic Center. The Carina-Sagittarius and Scutum-Crux 
   arms are labeled to make the figure easy to read. The three common reddening groups identified in Fig.~10 are
   here drawn with the same color-coding. }
  \end{figure}

\noindent
A second component is associated with the HII regions
RCW~83 (311.85, -00.54) and RCW~85 (313.40, -00.36)  and 
coincides with the Carina-Sagittarius arm at a distance of
$\sim$1.5-2.0 kpc, where NGC~5617 is located, together with Lynga~1 and 2, Hogg~17, Trumpler~22
and NGC~5606. This component possesses a velocity in the range -24 to -31 km/sec, very close
to the $\sim$-35.77$\pm$ 0.82 km/sec recently measured by Mermilliod et al (2008) for NGC~5617. 
In fact, if we convert this radial velocity value into V$_{lsr}$  adopting Russeil (2003) values
for the solar motion (U,V,W)=(-10.4,14.8,7.3) km/sec, we obtain -32.3 km/sec.
This similarity in velocity
strengthens the spatial coincidence of this component,
which is clearly associated with the main body of Carina-Sagittarius arm (Vall\'ee 2005, Russeil 2003)
at this longitude.\\
Our analysis in Sect.~10 fully agrees with this scenario, since we detected -together with NGC~5617- 
young field stars at heliocentric distance
between 1.5 and 3.0 kpc, which encompass the whole width of the Carina spiral arm (Bronfman et al. 2000).\\

\noindent
Finally,  a third group is detected behind CRW~83 and 85, with velocity ranging from -46 to -52
km/sec. Russeil et al. (1998) assign to this group a distance of 3.4$\pm$0.9 kpc, which
would place it beyond the Carina-Sagittarius arm.\\
Unfortunately, there is only a marginal overlap with the most distant group we found in Sect.~10,
which, according to our study, would lie further away . This non-coincidence
might also be due to the well known difficulties in deriving distances from radial velocities.\\

\noindent
In order to clarify whether this group is tracing a population belonging to a more distant arm,
we make use of the Vall\'ee (2005) spiral arm description of the Milky Way, as depicted in
Fig.~12.  Vall\'ee (2005) performed a statistical analysis of the data available in the literature on the properties
of the spiral arms in the Milky Way, like arm number, pitch angle, arms' shape and distance. He then provided
an analytical description to draw spiral arms using the most updated values of their parameters. 
We caution the reader that
this description we are going to use here is by no means an absolute one, since parameter values
changes as long as new data are being accumulated.\\

\noindent
We use the same color-coding as in Figs~10 and 11 to position in the helio-centric distance direction
the estimated location of the three groups we claim we have detected. The two closest groups, as anticipated in Sect.~10,
are suggested to likely trace the Carina-Sagittarius arm, at least as predicted by Vall\'ee (2005) model.\\

\noindent
The third group, although significantly more spread in distance, overlaps with the fourth quadrant portion of
the inner Galaxy Scutum-Crux arm at an heliocentric distance of about 5 kpc, and therefore we consider it a 
group of stars candidate to belong to this more distant arm. \\

\noindent
To lend further support to this scenario, we 
finally compare our distance determinations with the expected positions of the Carina-Sagittarius and Scutum-Crux
arm from another source, say the Russeil (2003) Galactic spiral arms description. 
In that study, a collection of data from different spiral arm indicators (H$_{\alpha}$, CO, radio continuum, 
absorption features, and H$_{109\alpha}$)   
is performed, and an analytical fitting is done on the data to predict the number and mean position of the various
arms in the plane of the disk. Russeil (2003) concludes that the data are best fitted by a four-arm model,
where, in the direction we are considering here, the Carina-Sagittarius arm is at about 2 kpc from the Sun,
and the Scutum-Crux arm at about 5 kpc from the Sun, in agreement with Vall\'ee (2005) statistical models.\\
\noindent
This seems to reinforce our results that the most distant group we detect represents a population
candidate  to be part of the  Scutum-Crux arm. \\

\noindent
To close, we stress that the results presented in this study have to be considered preliminary, and, in this
respect, a spectroscopic follow-up of these stars would be very welcome to confirm or deny our findings.

\begin{acknowledgements}
This study is based on data acquired at Las Campanas Observatory. The author is very grateful
for the technical support and the kind environment at the Observatory. The author also
acknowlegdes very fruitful exchange of information with Cameron Reed, Ruben A. V\'azquez, 
and Brian Skiff. The crucial help of an anonymous referee in improving the paper presentation is
also deeply recognized. 
We finally made use of the WEBDA database, maintained at Vienna 
University by E. Paunzen, and the SIMBAD database. 
\end{acknowledgements}

\end{document}